\documentclass[5p,times]{elsarticle}
\usepackage{amssymb,amsmath}
\usepackage[english]{babel}
\usepackage{graphicx}
\usepackage{color}

\journal{Computer Physics Communication}

\begin{document}

\begin{frontmatter}
  \title{Convergence of Stochastic Approximation Monte Carlo and\\ modified Wang-Landau algorithms: Tests for the Ising model}
  \author[]{Simon Schneider\corref{cor1}}
  \ead{Simon.Schneider@itp.uni-leipzig.de}
  \author[]{Marco Mueller\corref{cor2}}
  \ead{Marco.Mueller@itp.uni-leipzig.de}
  \author[]{Wolfhard Janke\corref{cor3}}
  \ead{Wolfhard.Janke@itp.uni-leipzig.de}
  \address{Institut f\"ur Theoretische Physik, Universit\"at Leipzig, Postfach 100920, D--04009 Leipzig, Germany}

  \begin{abstract}

  We investigate the behavior of the deviation of the estimator for the density
  of states (DOS) with respect to the exact solution in the course of Wang-Landau
  and Stochastic Approximation Monte Carlo (SAMC) simulations of the
  two-dimensional Ising model. We find that the deviation saturates in the
  Wang-Landau case. This can be cured by adjusting the refinement scheme. To this
  end, the $1/t$-modification of the Wang-Landau algorithm has been suggested. A
  similar choice of refinement scheme is employed in the SAMC algorithm. The
  convergence behavior of all three algorithms is examined. It turns out that the
  convergence of the SAMC algorithm is very sensitive to the onset of the
  refinement. Finally, the internal energy and specific heat of the  Ising model
  are calculated from the SAMC DOS and compared to exact values. 

  \end{abstract}


\end{frontmatter}

\section{Introduction}
\label{sec:introduction}

The Wang-Landau algorithm~\cite{WangLandau,wang2001determining} has proven to
be a very efficient tool for determining the density of states (DOS) of
statistical systems near phase transitions where traditional local importance
sampling algorithms like the Metropolis algorithm are likely to run into
critical slowing down or become trapped in local free-energy
minima~\cite{janke2016thermodynamics}. It has, however, been pointed out that
the error of the estimator for the DOS obtained by the Wang-Landau algorithm
cannot be made arbitrarily small just by using longer
simulations~\cite{yan2003fast}, the (systematic) error saturates at some
(small) value. To overcome this, it has been suggested to change the behavior
of the refinement parameter in the $1/t$ modification ($1/t$-WL) of the
Wang-Landau algorithm in order to circumvent the error
saturation~\cite{BelPerChem,BelPerPhysRev,zhou2008optimal}.

Another approach is the Stochastic Approximation Monte Carlo (SAMC) algorithm
first introduced in Ref.~\cite{liang2006theory} and refined in
Ref.~\cite{SAMC}, which works similar to the modified Wang-Landau algorithm
regarding the choice of refinement scheme. While the algorithm proposed by
Belardinelli and Pereyra has been tested for the Ising model~\cite{BelPerChem,
BelPerPhysRev}, for the calculation of multidimensional
integrals~\cite{belardinelli2008analysis} and was applied to lattice polymer
models~\cite{swetnam2011improving}, the SAMC algorithm has only been tested for
an artificial, non-physical model with a very small number of
states~\cite{SAMC} compared to models currently studied in statistical physics
and for an off-lattice polymer model~\cite{SAMCpoly}. 

The standard test case, the Ising model, however, is still missing. After
summarizing a few basic properties of this paradigm model needed for our
purposes in the next section, we discuss briefly the behavior of the
Wang-Landau and $1/t$-WL algorithms in Sections~\ref{sec:saturation}
and~\ref{sec:1_t}, confirming the results of Belardinelli and Pereyra. In
Section~\ref{sec:samc} we describe the SAMC algorithm and in
Section~\ref{sec:error} we present our results on the accuracy of the SAMC
algorithm applied to the two-dimensional Ising model for long simulation times.
Finally, Section~\ref{sec:conclusion} contains our conclusions.

\section{The Ising model}
\label{sec:ising}

First introduced in 1D by Ernst Ising in 1925 in his doctoral
thesis~\cite{ising}, the generalisation of the Ising model to higher
dimensions, especially 2D and 3D, has become the ``drosophila of computational
physics''. Despite its conceptual simplicity it shows nontrivial behavior
already in the 2D case like a temperature-driven continuous phase transition at
finite temperature. The model consists of a lattice of spins $s_i$ which can
take the values $s_i = \pm 1$. The Ising Hamiltonian reads
\begin{equation}
H = - J \sum\limits_{\langle i,j\rangle}^{}s_i s_j - h \sum\limits_{i}^{}s_i \;,
\label{eg:ising}
\end{equation}
where $\langle i,j\rangle$ denotes a sum only over nearest-neighbour spins, the
coupling constant $J$ sets the energy scale and $h$ is an external magnetic
field. The partition function of the 2D Ising model with periodic boundary
conditions was found by Lars Onsager in the infinite-volume limit for the case
$h=0$ in 1944~\cite{onsager} and extended to finite square lattices by Kaufman
in 1949~\cite{kaufman}. An easy way to compute the DOS for finite lattices was
only found much later by Beale~\cite{beale} using the fact that it is possible
to rewrite the partition function $Z$ in the following way:
\begin{equation}
 Z = \sum_{\mu}e^{-\beta H} = \sum_E g(E) e^{-\beta E} \;,
\end{equation}
where $\beta=1/k_\mathrm{B} T$ and $g(E)$ denotes the number of microstates with energy
$E$. The first sum is over all possible microstates $\mu$ while the second is
carried out over all energies available to the system. The quantity $g(E)$ is
identical to the density of states. The results of Beale will be used in the
following to compare the performances of the SAMC and Wang-Landau algorithms.

\section{Error saturation in the Wang-Landau algorithm}
\label{sec:saturation}

\begin{figure*}[t]
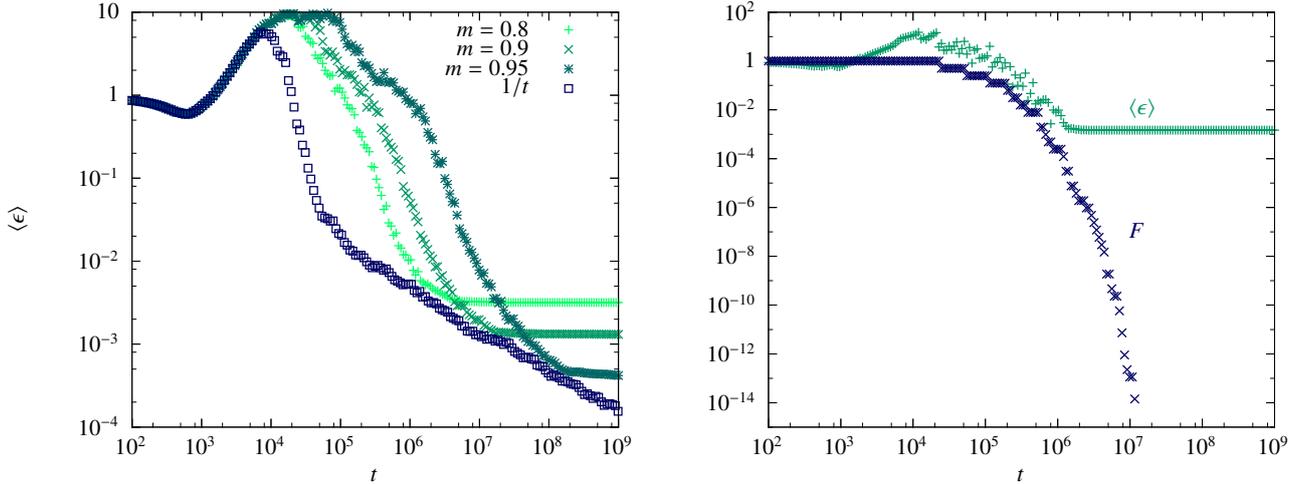

  \centering
  \fontsize{9}{0}
  \scalebox{.95}{\input{wl_saturation.texi}}
  \scalebox{.95}{\input{wl_refinement.texi}}
  \caption{(left) The behavior of the average deviation from the exact solution
    $\langle \epsilon(t) \rangle_E$ for the $8\times 8$ Ising model for the
    Wang-Landau and $1/t$-WL algorithm over attempted spin flips (MC time) $t$
    for different flatness parameters $m$. The data was obtained by averaging
    over 40 independent runs of the algorithm to reduce statistical noise.
    (right) The behavior of the average deviation from the exact solution
    $\langle \epsilon(t) \rangle_E$ and the logarithm $F$ of the refinement
    parameter over MC time $t$ for one single simulation with
    flatness parameter $m = 0.8$ with extended $y$-range to show the behavior
    of $F$.}
  \label{fig:wl_m}
\end{figure*} 
The Wang-Landau algorithm attempts to sample the density of states $g(E)$ by
ensuring that during the simulation each energy is visited equally often. This
is equivalent to sampling with probability proportional to $g^{-1}(E)$ .
Because the DOS is \emph{a-priori} unknown, this is done by starting with a
very rough estimator $\widehat{g}(E)$ for $g(E)$, usually $\widehat{g}(E) = 1$
for all values of $E$. One also needs to keep track of the histogram $H(E)$,
which stores the number of times the states with energy $E$ are visited during
the simulation. The density of states $\widehat{g}(E)$ is then refined using a
parameter $f$, usually set to $f = e = 2.71828...$ initially. The refining
process works as follows:
\begin{enumerate}
  \item Pick a starting state $\mu$ and calculate its energy $E_\mu$. Set
    $H(E)=0$ and $\widehat{g}(E) = 1$  for all energies  
  \item Propose a randomly chosen update to get a new state $\nu$ with energy
    $E_\nu$\label{wl2}
	\item Calculate $\widehat{g}(E_\mu)/\widehat{g}(E_\nu)$\label{wl3}
	\begin{enumerate}
    \item if $\widehat{g}(E_\mu)/\widehat{g}(E_\nu)\geq1$ accept the update and
      set \\$\widehat{g}(E_\nu)\rightarrow \widehat{g}(E_\nu)\cdot f$ and\\
      $H(E_\nu)\rightarrow H(E_\nu)+1$
    \item if $\widehat{g}(E_\mu)/\widehat{g}(E_\nu)<1$ accept the update with 
      probability $\widehat{g}(E_\mu)/\widehat{g}(E_\nu)$  
		\begin{enumerate}
      \item If accepted, set \\$\widehat{g}(E_\nu)\rightarrow
        \widehat{g}(E_\nu)\cdot f$ and\\ $H(E_\nu)\rightarrow H(E_\nu)+1$
      \item If rejected, set \\$\widehat{g}(E_\mu)\rightarrow
        \widehat{g}(E_\mu)\cdot f$ and\\ $H(E_\mu)\rightarrow H(E_\mu)+1$
		\end{enumerate}
	\end{enumerate}
  \item After a suitable number (e.g. 1000) of repetitions of steps
    (\ref{wl2}-\ref{wl3}), check the histogram for flatness.
	\begin{enumerate}
    \item If the histogram is not flat, go back to step \ref{wl2}
    \item If the histogram is flat, reset $H(E)$ to $0$ for all values of $E$,
      adjust $f \rightarrow f^{\frac{1}{2}}$, then go  to step \ref{wl5}
	\end{enumerate}
  \item Check whether $f$ is smaller than some predefined final value
    $f_\text{final}$.\label{wl5}
	\begin{enumerate}
    \item If $f$ is not smaller than $f_\text{final}$, go back to step
      \ref{wl2}
    \item If $f$ is smaller than $f_\text{final}$, the calculation is finished
      and $\widehat{g}(E)$ obtained during the simulation is an estimator for
      the real density of states
	\end{enumerate}
\end{enumerate}
We use a slight variation of the algorithm: Since our goal is to reproduce the
error saturation, we do not terminate the algorithm at a certain value
$f_\text{final}$, but run it as long until we observe the saturation of the
error.  It is convenient to work with the logarithms of the values $S :=
\ln\widehat{g}$ and $F := \ln f$ in order to fit the large numbers into double
precision variables and to replace the multiplication $\widehat{g}\cdot f$ by
an addition $S+F$. Physically, $S(E)$ is the (configurational) microcanonical
entropy (in units where $k_\mathrm{B}=1$)~\cite{Schierz2016}.
One sees that for long simulation times, the flatness criterion is satisfied in
every iteration of the algorithm, provided that the time between subsequent
flatness checks is long enough. Thus the refinement parameter $F$ is made
smaller in every iteration, which means that it decreases geometrically. This
has been shown to prevent the convergence of the error to
zero~\cite{BelPerChem}.  A somewhat subtle point in the Wang-Landau algorithm
is the notion of flatness. Here we use the criterion that the histogram is
flat, if for all energies $E$, $H(E)$ falls into the interval
$\left[m\overline{H},(2-m)\overline{H}\right]$, where $\overline{H}$ denotes
the average value of the histogram and $m$ is the so called flatness parameter,
usually chosen between $0.7$ and $1$. The flatness parameter, however, cannot
overcome the fundamental saturation problem of the Wang-Landau algorithm, the
saturation just sets in at a lower deviation value for higher
$m$~\cite{BelPerChem}. On the left of Fig.~\ref{fig:wl_m} the behavior of the
average deviation from the exact solution,
\begin{equation}
  \label{eq:error}
  \langle \epsilon(t) \rangle_E = \frac{1}{N_E-1}\sum_E\left|\frac{S(E,t)-S_{\text{Beale}}(E)}{S_{\text{Beale}}(E)}\right| \;,
\end{equation}
during Wang-Landau simulations of the $8\times 8$ Ising model is shown. Here,
we have introduced the Monte Carlo (MC) time $t$ in units of  updates
in the sense of step 2 of the Wang-Landau algorithm,  $N_E$ denotes the number
of different energy values and $S_{\text{Beale}}$ is the exact value of $S$
from the Beale solution~\cite{beale}. In case of the Ising model, the update is
the spin flip of a uniformly randomly chosen spin. The estimator $S(E,t)$ is
normalized with respect to the exact DOS of the ground states.  It is visible
that the deviation saturates, even for the strictest choice of flatness
parameter, reproducing the results of Belardinelli and
Pereyra~\cite{BelPerChem}. On the right of Fig.\ \ref{fig:wl_m} the behavior of
the logarithm $F$ of the refinement parameter $f$ during the saturation phase
is depicted. From the beginning of the saturation phase, it decreases too fast
to correct the estimator for the DOS significantly.

\section{The 1/t-modification of the Wang-Landau algorithm}
\label{sec:1_t}

The error saturation can be corrected by appropriately adapting the refinement
parameter in order to prevent its fast decrease. This was suggested by
Belardinelli and Pereyra in Ref.~\cite{BelPerPhysRev}. Their algorithm avoids
the notion of flatness of the histogram and, in the first phase, only checks
whether all energies have been visited and refined at least once by the current
$F_t$. If $F_t \leq N_E/t$, where $N_E$ denotes the number of energies
available to the system, the algorithm enters its second phase, the histogram
is no longer checked and $F_t$ is updated each simulation step $t$ according to
$F_t=N_E/t$. For long times, $F_t$ now converges much slower towards zero than
in the original Wang-Landau case. This can avoid the saturation of the
deviation as is shown in Fig.\ \ref{fig:wl_m}. The deviation for the
$1/t$-algorithm does not saturate in the simulation time of $10^9$ attempted
spin flips for the $8\times8$ Ising model. Another limitation of the
Wang-Landau algorithm, which is the fact that the energy range of the model
needs to be known beforehand is still present in this algorithm: One needs to
know all admissible energies of a system in order to discern between histogram
bins that have not been visited by the algorithm yet, but are available to the
system in principle, and histogram bins that can never be visited because the
system can not access those energies. This is not a big problem in the Ising
model, because the energy range is known here, but can be a problem for systems
with unknown ground states like spin glasses or lattice
polymers. An extension to the algorithm which can
overcome this problem has been proposed in Refs.~\cite{wuest2009versatile,swetnam2011improving}.
\begin{figure*}[t]
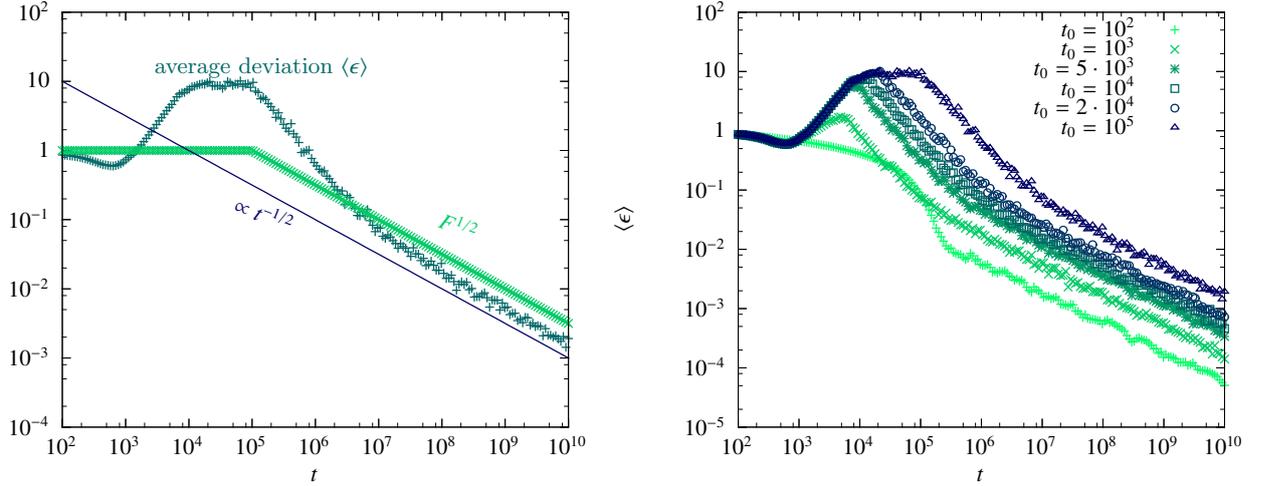

  \centering
  \fontsize{9}{0}
  \scalebox{.95}{\input{samc_8_behavior.texi}}
  \scalebox{.95}{\input{samc_8.texi}}
  \caption{
    (left) The behavior of the average deviation from the exact solution
    $\langle \epsilon(t) \rangle_E$ over MC time $t$ for the SAMC
    simulation of the $8\times 8$ Ising model (pluses) and of the square root
    $F^{1/2}$ of the refinement parameter $F$ (crosses). $t_0$ was chosen as
    $t_0 = 10^5$ here. The straight line is the graph of $t^{-1/2}$. This line is
    roughly parallel to the graph of the deviation. 
    (right) The behavior of the average deviation from the exact solution for
    different choices of $t_0$. All data is averaged over 40 independent
    simulation runs in order to reduce statistical errors.}
  \label{fig:samc}
\end{figure*}

\section{The SAMC algorithm}
\label{sec:samc}

The SAMC algorithm~\cite{liang2006theory,SAMC} is an algorithm which is based
on stochastic approximation algorithms~\cite{stochapp} and is similar to the
Wang-Landau algorithm as it tries to obtain an estimator for the DOS by using a
refinement parameter. The obtained estimator for the DOS is proven to converge
to the real DOS almost surely if the sequence of refinement parameters $F_t$
satisfies
\begin{align}
  \sum_{t=1}^{\infty}F_t = \infty \quad \text{ and} \quad
  \sum_{t=1}^{\infty}F_t^\zeta < \infty \;.
\end{align} 
for some $\zeta \in (1,2)$ along with necessary conditions on the proposal
distribution~\cite{liang2006theory,SAMC}. The Wang-Landau algorithm violates
this criterion, as the sequence $F_t^{\text{WL}}$ is proportional to ${2^{-t}}$
for large $t$ and $\sum_{t=1}^{\infty}{2^{-t}} < \infty$ for Wang-Landau. The
SAMC algorithm uses the sequence
\begin{equation}
  F_t^{\text{SAMC}}=\frac{t_0}{\max(t_0,t)}
\end{equation}
instead, where $t_0 > 1$ is an arbitrary value which can be chosen according to
the problem. This sequence fulfils both criteria because it behaves like $1/t$
for long simulation times and the harmonic series $\sum_{t=1}^{\infty}1/t$
diverges, while the series $\sum_{t=1}^{\infty}1/t^{\zeta}$ converges for
$\zeta > 1$. This is very similar to the $1/t$-WL case, where $F$ exhibits the
same long-time behavior. Following Liang, Liu, and Carroll~\cite{SAMC}, one
special case of their algorithm that aims for a flat histogram works as follows:
\begin{enumerate}
  \item Pick a starting state $\mu$ and calculate its energy $E_\mu$. Set
    $S(E)=0$ for all energies  
  \item Randomly propose an update to get a new state $\nu$ with energy
    $E_\nu$\label{samc2}
  \item Calculate $e^{[S(E_\mu)-S(E_\nu)]}$\label{samcl3}
  \begin{enumerate}
    \item if $e^{[S(E_\mu)-S(E_\nu)]}\geq1$ accept the update and set
      $S(E_\nu)\rightarrow S(E_\nu)+F_t$ 
    \item if $e^{[S(E_\mu)-S(E_\nu)]}<1$ accept the update with a probability
      $e^{[S(E_\mu)-S(E_\nu)]}$  
    \begin{enumerate}
      \item If accepted, set $S(E_\nu)\rightarrow S(E_\nu)+F_t$  
      \item If rejected, set $S(E_\mu)\rightarrow S(E_\mu)+F_t$  
    \end{enumerate}
  \end{enumerate}
  \item Go back to step 2, repeat the algorithm until $F_t$ is smaller than some
    predefined value.\label{samc5}
\end{enumerate} 
One sees that the behavior of $F_t^{\text{SAMC}}$ is, in contrast to the
Wang-Landau algorithm, fully deterministic, so the runtime of the algorithm can
be predicted. Furthermore the SAMC algorithm avoids the checks for histogram
flatness one has to do while using Wang-Landau by simply starting the
decreasing behavior of $F_t$ at some defined time. The general algorithm provides
the possibility to divide the energy space into subregions in order to perform
a biased sampling~\cite{liang2006theory,SAMC}, which is not considered here, as
the energy landscape of the Ising model is uncomplicated enough. Nevertheless,
it has been noted before that sampling with other weight distributions than the
inverse DOS can improve the quality of estimators obtained by Monte Carlo
simulations~\cite{hesselbo1995monte}. An example of the impact of biased SAMC
sampling is described in Ref.~\cite{SAMCpoly}. It should also be noted that
none of the algorithms presented here can be viewed as a special case of
Metropolis sampling with sampling distribution $g^{-1}(E)$ since the weights
are perpetually modified. Therefore detailed balance is violated and the
sampling process is non-Markovian.

\section{Simulating the Ising model with SAMC}
\label{sec:error}
\begin{figure}[t]
  \centering
  \fontsize{9}{0}
  \scalebox{.95}{\input{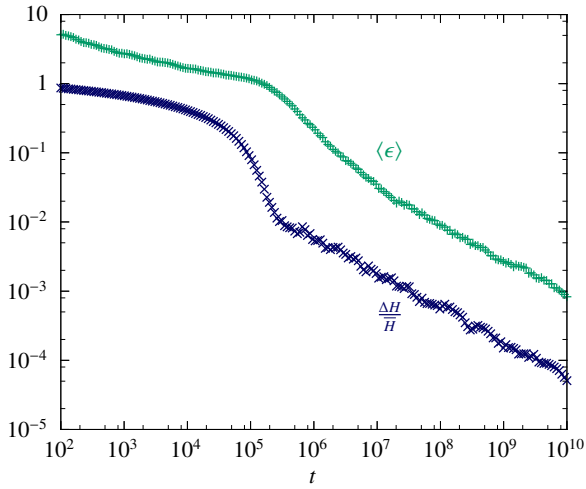}}
  \caption{
    Green pluses: The behavior of the average deviation from the exact solution
    $\langle \epsilon(t) \rangle_E$ over MC time $t$ for the SAMC
    simulation of the $8\times 8$ Ising model. $t_0$ was chosen as $t_0 = 100$
    here.
    Blue crosses: The behavior of $\Delta H /\overline{H}$ over MC time $t$.
    The data was obtained by averaging over 40 independent runs of the algorithm
    to reduce statistical noise.}
  \label{fig:samc_100_flat}
\end{figure}
\begin{figure*}[t]
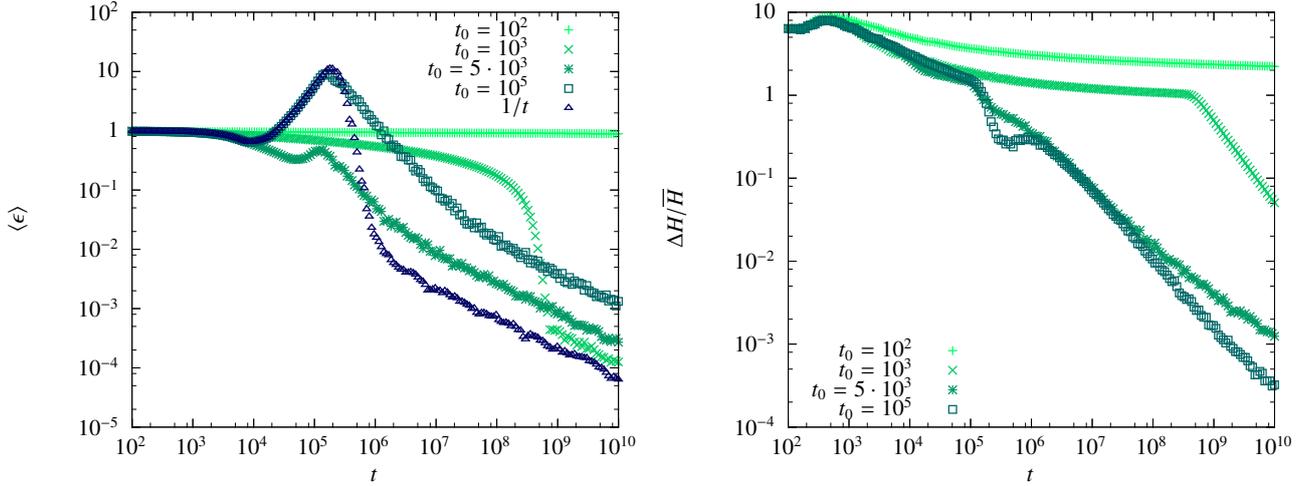

  \centering
  \fontsize{9}{0}
  \scalebox{.95}{\input{samc_16.texi}}
  \scalebox{.95}{\input{samc_16_flatness.texi}}
  \caption{
    (left) The behavior of the average deviation from the exact solution
    $\langle \epsilon(t) \rangle_E$ over MC time $t$ for SAMC for
    different $t_0$ and $1/t$-WL simulations of the $16\times 16$ Ising model.
    (right) The flatness of the histogram over MC time $t$ for
    different choices of $t_0$. The data was obtained by averaging over 40
    independent runs of the algorithm to reduce statistical noise.}
  \label{fig:samc16}
\end{figure*} 

In this section we show the results of simulations of the Ising model for
various lattice sizes using SAMC. The Hamiltonian for the Ising model without
an external magnetic field is the special case $h=0$ of the Hamiltonian defined
in Eq.\ (\ref{eg:ising}).
The deviation for the DOS with respect to the Beale solution is calculated
according to Eq.\ (\ref{eq:error}) where $\widehat{g}(E)$ is normalized with
respect to the exact DOS of the ground states. All simulations were started at
one of the ground states of the Ising model, we chose all spins equal to $+1$.
All data was obtained by averaging over 40 runs of the respective algorithms
with independent seeds of the random number generator. Due to the symmetry of
the density of states of the Ising model only the ferromagnetic states with
energies from $E = E_0 = -2L^2$ up to $E = 0$ were considered.

\subsection{The $8\times 8$ Ising model}

In Fig.\ \ref{fig:samc} on the left side the behavior of the deviation from the
exact solution and the refinement parameter during the run of the simulation
for the $8\times 8$ Ising model with $t_0 = 10^5$ is shown. It is visible that,
in the beginning, the deviation increases, which is due to the rough refinement
with $F_t = 1$. From the onset of the $F_t \propto 1/t$ behavior at $t = t_0 =
10^5$ however, the deviation steadily decreases until the end of the simulation
which was even longer than the Wang-Landau simulations in Section
\ref{sec:saturation}. For later times, the deviation is proportional to
$F^{1/2} \propto t^{-1/2}$, as indicated by the parallel lines in the
double-logarithmic plot. This is similar to the findings of Belardinelli and
Pereyra on their $1/t$-WL algorithm~\cite{BelPerPhysRev}. The attained
deviation after $10^9$  updates is comparable to  our \emph{least
accurate} estimator from the Wang-Landau simulation with $m=0.8$. This can be
improved by choosing a more appropriate $t_0$. One sees that in the time
between $10^4$ and $10^5$  updates the deviation stays at a constant,
large value.  In this phase the rough refinement with $F_t=1$ explores the energy
landscape and reaches all available energies to ensure that the refinement can
take place properly in the next stage. This does not necessarily generalize to
more complicated models like lattice polymers, where some ``hidden'' (low)
energies can only be visited when the estimator of the DOS is already
sufficiently accurate~\cite{wuest2012optimized}. In Fig.\ \ref{fig:samc} on the
right side, the behavior of the deviation for SAMC simulation of the $8\times
8$ Ising model is shown for various values of $t_0$ ranging from $t_0 = 100$ to
$t_0 = 10^5$. For the large values of $t_0$, the plateau phase is visible. For
$t_0 = 100$, however, we see a decline in the deviation from the beginning,
making it the most efficient choice for achieving small errors at the end of
the simulation. This seems to be due to the very small phase space of the
$8\times 8$ Ising model.  To investigate this, we need to take a look at how
flat the histogram gets during the simulation. If the histogram never really
gets flat, this is an indicator that the simulation is not able to sample the
complete energy space, because some states are never visited. To this end, we
take a look at the parameter $\Delta H /\overline{H}$, where $\Delta H$ is the
difference between the largest and the smallest value of $H(E)$ and
$\overline{H}$ is the average of $H(E)$ over all energies. This parameter can
be used as an indicator for the flatness of the histogram because it measures
the ratio of the largest deviations to the average. For a perfectly flat
histogram, it is zero, while it is of the order of one if some energy values
are never visited. In Fig.\ \ref{fig:samc_100_flat}, this parameter is depicted
along with the deviation for the run with $t_0 = 100$. For the $8\times 8$
Ising model, the histogram becomes flat even for $t_0 = 100$, and $\Delta H
/\overline{H}$ goes as $t^{-1/2}$. 

\subsection{The $16\times 16$ Ising model}
\label{sec:error16}

\begin{figure}[h]
  \centering
  \fontsize{9}{0}
  \scalebox{.95}{\input{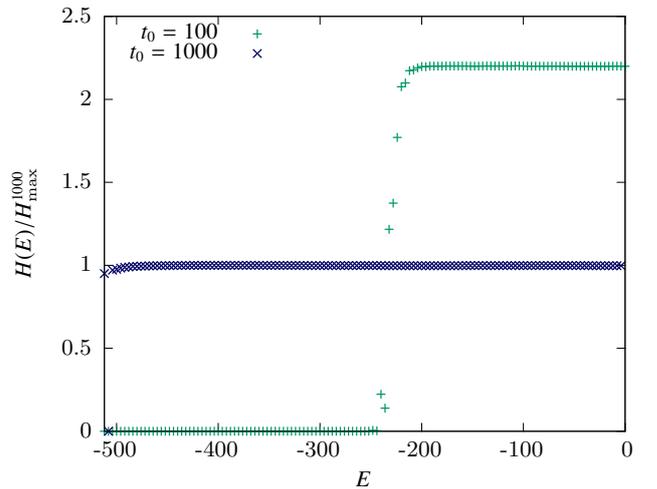}}
  \caption{
    Green pluses: The histogram after $10^{10}$ SAMC updates for the $t_0 =
    100$ case. Blue crosses: The histogram after $10^{10}$ SAMC updates for the
    $t_0 = 1000$ case. Both histograms are normalised to the maximum value of the
    $t_0 = 1000$ histogram. The histogram in the $t_0 = 100$ case is clearly not
    flat, the low-energy regions are undersampled.}
  \label{fig:samc_hist}
\end{figure} 
The impact of the small size of the energy space of the $8\times 8$ Ising model
becomes already clear by considering the $16\times 16$ Ising model. In Fig.\
\ref{fig:samc16}, the behavior of the deviation from the exact solution of the
DOS obtained by the SAMC algorithm for the $16\times 16$ Ising model is
depicted for different $t_0$ on the left. The flatness of the histogram for
those values of $t_0$ is shown on the right. For $t_0 = 100$, the algorithm
fails to generate a flat histogram in the simulation time of $10^{10}$ update
attempts.  The final histogram of that simulation is shown in Fig.\
\ref{fig:samc_hist}. The states below $E \approx -200$ are almost never
visited.
\begin{figure*}[t]
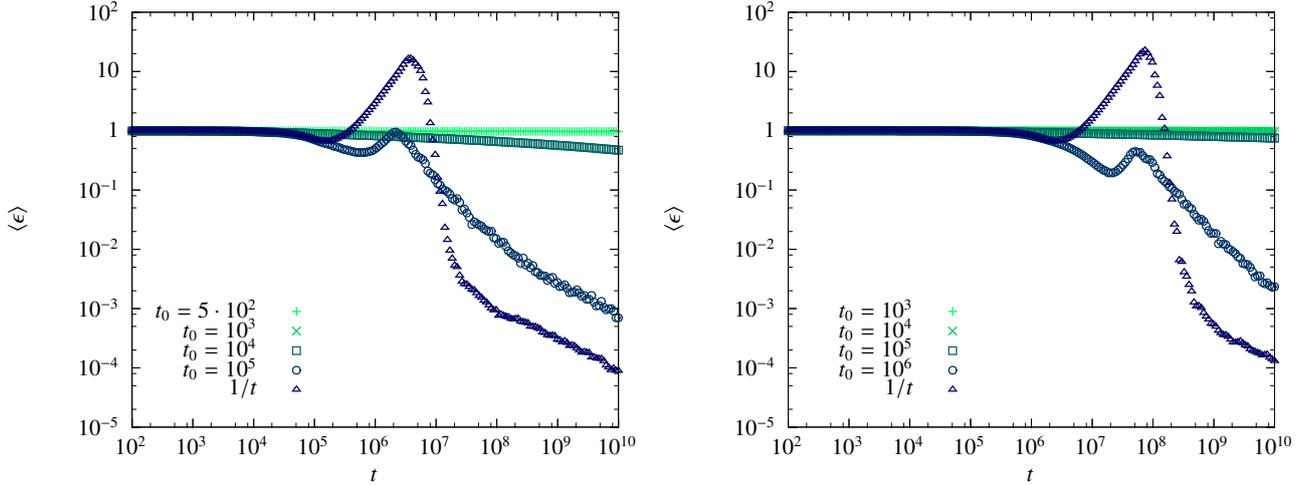

  \centering
  \fontsize{9}{0}
  \scalebox{.95}{\input{samc_32.texi}}
  \scalebox{.95}{\input{samc_64.texi}}
  \caption{
    The behavior of the average deviation from the exact solution $\langle
    \epsilon(t) \rangle_E$ over MC time $t$ for SAMC for different
    $t_0$ and $1/t$-WL simulations of the $32\times 32$ (left) and $64\times
    64$ (right) Ising model. The $1/t$-WL simulations converge faster than all
    SAMC simulations in both cases.}
  \label{fig:samc32}
\end{figure*}
\begin{figure}[h]
  \centering
  \fontsize{9}{0}
  \input{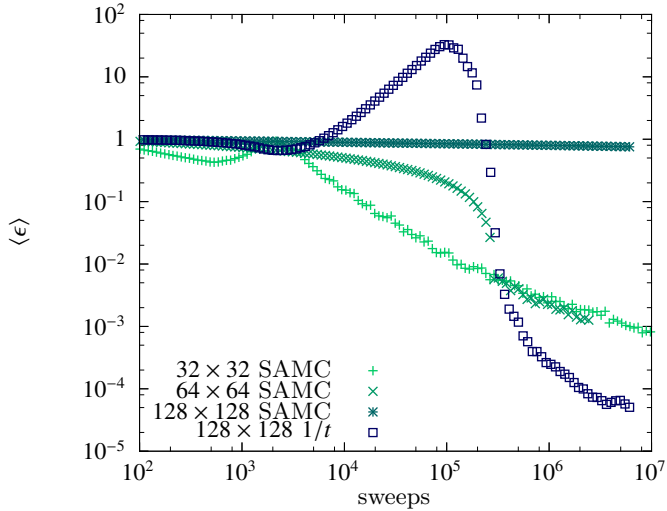}
  \caption{The behavior of the deviation $\langle \epsilon \rangle$ of the
    simulated DOSes from SAMC and $1/t$-WL simulations for various (larger)
    lattice sizes. $t_0$ was chosen as $t_0 = 100L^2$ for all SAMC simulations
    in order to check the rule of thumb given by Liang et al.~\cite{SAMC}. The
    time axis is given in sweeps ($1 \text{ sweep} \equiv L^2$ updates)
    here to allow a comparison between the different lattice sizes.}
  \label{fig:samc_sweeps}
\end{figure} 
This is also reflected in the DOS failing to converge, the deviation stays
roughly at one for the whole simulation time. For $t_0 = 1000$ the behavior
looks similar to that for $t_0 = 100$ in the $8\times8$ Ising case, first
converging slowly, then, after a steep decline of the deviation, transitioning
into the $\langle \epsilon \rangle \propto t^{-1/2}$ behavior. For higher
$t_0$, the histogram steadily flattens, then converges in the $\langle \epsilon
\rangle \propto t^{-1/2}$ manner. This shows that the choice of an appropriate
$t_0$ is crucial. In their original work~\cite{SAMC}, Liang et al. state that a
rule of thumb for choosing an appropriate $t_0$ is between $2N_E$ and $100N_E$,
where $N_E$ is the number of energy subregions. In our case, this is equal to
the number of histogram bins. For the $16\times16$ Ising model, we have
$N_E=128$. Our finding that $t_0 = 100$ is too small, whereas $t_0 = 1000$
suffices, is in good agreement with this rule. The problem of choosing $t_0$
appropriately is avoided in the $1/t$-WL algorithm. This algorithm determines
the starting point of the $1/t$-behavior by checking the sequence $F_t$ itself
by comparing it to the desired $F_t \propto 1/t$ sequence. This procedure is
more self-contained than the procedure used by SAMC, but has to rely on
stochastic quantities (here the number of visits in each histogram bin) and an
\emph{a-priori} knowledge of the number of accessible energies which may be a
problem for more complicated
models~\cite{wuest2009versatile,swetnam2011improving,wuest2012optimized}.
This makes it hard to predict the runtime of the algorithm.  The convergence of
the $1/t$-algorithm in this example seems to be very good, even exceeding our
best choice of $t_0$ for SAMC. This is visible in the left of Fig.\
\ref{fig:samc16}.  \subsection{Larger lattice sizes} For larger lattice sizes
the necessity of choosing an appropriate value for $t_0$ becomes even more
apparent. This is visible in Fig.\ \ref{fig:samc32}, where the deviation from
the exact solution during SAMC and $1/t$-WL simulations of the $32\times32$ and
$64\times64$ Ising model is depicted. As expected, the time needed to explore
the energy landscape increases with the lattice size (which is, of course,
equivalent to increasing phase space size).  If $t_0$ is chosen too small in
the SAMC simulations, the algorithm fails to converge in our simulation time of
$10^{10}$ updates. The $1/t$-WL algorithm, on the other hand, adapts to this
increase by its flatness criterion which forces the algorithm to take all
energies into account. This is visible in the figure by a rapid decrease of
$\langle \epsilon \rangle$ when the $F \rightarrow F/2$ stage of the algorithm
starts. Finally, the $F_t \propto 1/t$ behavior sets in and the deviation $\langle \epsilon
\rangle$ decreases like $t^{-1/2}$.  While the $1/t$-WL algorithm manages to
find the right time for the onset of the $1/t$-stage by itself in case of the
Ising model, for SAMC it is up to the user to decide which $t_0$ is
sufficiently large to ensure convergence. Our simulations for the quite
uncomplicated Ising model already give a hint that this might be an intricate
task for more difficult models. In Fig.\ \ref{fig:samc_sweeps}, we show that
the rule of thumb mentioned in Section~\ref{sec:error16} is no longer
applicable for the $128\times 128$ Ising model: All SAMC simulations shown in
this figure were carried out using $t_0 = 100L^2$, which is even larger than
the value of $100N_E$ given by the rule of thumb. While still converging for
the $32\times 32$ and $64 \times 64$ models, the algorithm fails to reduce
$\langle \epsilon \rangle$ within $10^7$ sweeps for the $128 \times 128$ model.
Additionally, we show the performance of the $1/t$-WL algorithm for the $128
\times 128$ Ising model, which again outperforms the SAMC variants in terms of
faster convergence. 

\subsection{Thermodynamic quantities of the Ising model obtained by SAMC} 
\label{sec:samcising}

\begin{figure*}[t]
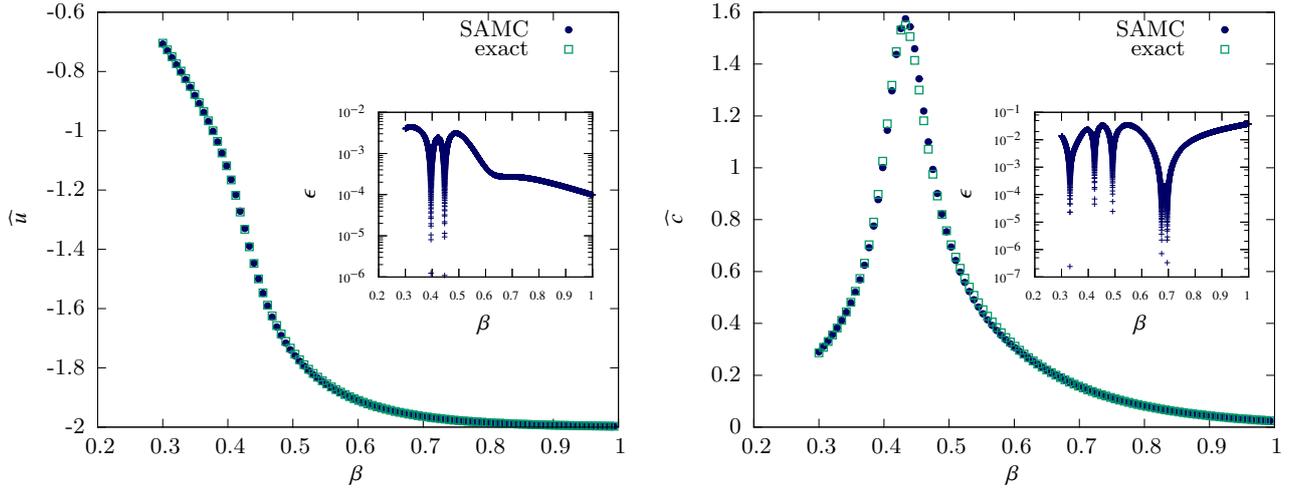

  \centering
  \fontsize{9}{0}
  \scalebox{.95}{\input{samc_16_reweight_e_hat.texi}}
  \scalebox{.95}{\input{samc_16_reweight_c_hat.texi}}
  \caption{
    The internal energy $u$ per spin (left) and the specific heat $c$ per spin
    (right) for the $16 \times 16$ Ising model depending on inverse temperature
    $\beta$ compared to data obtained from the exact Beale solution. The two
    curves coincide. The insets show the relative deviations $\epsilon$ of the
    internal energy and specific heat from the exact solution. }
  \label{fig:samc_16}
\end{figure*} 
The estimators for the DOS obtained in the SAMC simulation were used to
obtain an estimate of the internal energy $u$ and the specific heat $c$ per
lattice site of the $16 \times 16$ Ising model. The estimators for $u$ and $c$
were obtained by using the fact that for arbitrary inverse temperatures $\beta$
we can calculate $\widehat{u}(\beta) = \widehat{U}(\beta)/V$ and
$\widehat{c}(\beta) = \widehat{C}(\beta)/V$ from the DOS by
\begin{equation}
  \begin{aligned}
    \widehat{U}(\beta)&=\widehat{\langle E \rangle} = \frac{\sum_E E\;
      \widehat{g}(E) e^{-\beta E}}{\sum_E \widehat{g}(E) e^{-\beta E}} =
      \frac{\sum_E Ee^{S(E)-\beta E}}{\sum_E e^{S(E)-\beta E}} \;,\\
    \widehat{C}(\beta)&=\beta^2 (\widehat{\langle E^2 \rangle}
    -\widehat{\langle E \rangle}^2) = \beta^2\left(\frac{\sum_E E^2
      e^{S(E)-\beta E} }{\sum_E e^{S(E)-\beta E}}-\widehat{\langle  E \rangle}
      ^2 \right) \;.
  \end{aligned}
  \label{eq:u_reweight}
\end{equation}
In Fig.\ \ref{fig:samc_16} the curves obtained by this method are shown
compared to the exact curves from plugging Beale's solution into Eq.\
(\ref{eq:u_reweight}). The simulation parameters were $t_0 = 1000$ and the
total MC time was $10^{10}$. By the naked eye, for the internal energy no
differences can be spotted.  In the insets we show the relative deviations of
the two curves which is very small for the internal energy and small for the
specific heat. 

These results show that, if a run of SAMC has converged, its results can be
used for accurate calculations of thermodynamic properties of physical
models. Nevertheless, this procedure is not advisable for more complex systems,
because it does not allow a straightforward calculation of the errors of the
obtained physical quantities, since the error of the DOS is generally not
known. Therefore a
multicanonical~\cite{berg1991multicanonical,berg1992multicanonical,janke1992multicanonical}
production run using the inverse of the obtained estimator of the DOS as
weights is highly recommended. Moreover, a subsequent production run can single
out runs of the SAMC algorithm which have not converged. The data from
Section~\ref{sec:error} shows that this can be a serious issue even for the
Ising model, depending on the choice of $t_0$. 

\section{Conclusion}
\label{sec:conclusion}

The error saturation in the Wang-Landau algorithm found in
Refs.~\cite{BelPerChem} and \cite{BelPerPhysRev} was reproduced. A possible
cure for this is to use an alternative behavior of the refinement parameter as
is done in the $1/t$-WL and SAMC algorithms. The $1/t$-method inherits the
problem from the Wang-Landau algorithm of needing to know the range of
admissible energies for the considered model. In SAMC, this needs not to be
known beforehand, since histogram checking is not necessary in principle. The
SAMC algorithm, on the other hand, sometimes failed to converge in our examined
runtimes. This is caused by the simulation failing to explore the low-energy
states. Therefore no flat histogram can be produced. Since both variants of the
Wang-Landau algorithm regularly check the histogram for adequate flatness, it
is ensured that all energies are visited at least once. While the SAMC
algorithm should converge to the desired distribution if all conditions are
met, it is necessary to check if the histogram measured during the simulation
was really flat at the end.  This dampens the advantage of a predictable
runtime, since it is possible that a complete run of the algorithm turns out to
be unusable due to an inappropriate choice of $t_0$. Monitoring the flatness of
the histogram during the run is no help, because this would introduce a
stochastic quantity into the algorithm, making the runtime unpredictable and
require the same \emph{a-priori} knowledge of the admissible energy range as
the Wang-Landau algorithm and its modifications. The rule of thumb for the
choice of $t_0$ given by Liang et al.~\cite{SAMC} is violated even by the
$128\times128$ Ising model, showing that finding an appropriate $t_0$ can be a
quite cumbersome task. The $1/t$-WL algorithm suffers from a similar
restriction in this regard: While we could not find anything comparable in the
Ising model, other studies suggest that the overall convergence behavior can
also be sensitive to the details of the $1/t$-refining scheme for more
complicated models~\cite{zhou2008optimal,swetnam2011improving}. Regarding the
common features of the SAMC and $1/t$-WL algorithms, it seems reasonable to
assume that the proof of convergence for SAMC also extends to the $1/t$-WL
algorithm as well, since their long-time behavior is the same. Therefore the
choice of algorithm to apply to a certain problem is a practical one. With the
modifications proposed in Refs.~\cite{wuest2009versatile,
swetnam2011improving}, allowing it to adapt to \emph{a-priori} unknown energy
ranges, and in Refs.~\cite{zhou2008optimal, swetnam2011improving}, allowing it
to find the right time for the onset of the $1/t$-refinement, the $1/t$-WL
algorithm might be able to overcome its drawback for complicated systems with
unknown ground states. SAMC still has the advantage of allowing to generate
weights not only according to the density of states, but also according to
other distributions~\cite{liang2006theory,SAMC}, which can improve
estimators~\cite{hesselbo1995monte} and might prove useful for complex systems
like spin glasses or polymers, because sampling with distributions other than
the inverse density of states can speed up round trip times
significantly~\cite{trebst2004optimizing}.

\section*{Acknowledgments}

We thank B. Werlich, T. Shakirov, M. Taylor, and W. Paul for useful discussions
within the Halle/Leipzig Collaborative Research Center SFB/TRR 102, for which
we gratefully acknowledge funding by the Deutsche Forschungsgemeinschaft (DFG)
under project B04. This work has received further financial support by the
Deutsch-Franz\"osische Hochschule (DFH-UFA) through the Doctoral College
``${\mathbb L}^4$'' under Grant No.\ CDFA-02-07.

\section*{References}
\bibliographystyle{model1a-num-names}
\bibliography{paper}

\end{document}